\documentclass[superscriptaddress,letterpaper,twocolumn,aps,preprintnumbers,notitlepage,groupedaddress]{revtex4-1}
\usepackage{amsmath,amssymb,esint}
\usepackage{lipsum} 
\usepackage[latin9]{inputenc}
\usepackage{graphicx}
\usepackage{dcolumn}
\usepackage{bbold}
\usepackage{bm}
\usepackage[usenames]{xcolor}
\usepackage[colorlinks=true,linktocpage,bookmarks=true,citecolor=blue,linkcolor=blue,urlcolor=blue,hyperfootnotes=true,pageanchor=true]{hyperref}
\usepackage[caption=false,labelformat=simple]{subfig}
\usepackage{accents}
\usepackage[export]{adjustbox}
\usepackage{mathtools}
\usepackage{float}
\usepackage{mathrsfs}

\begin{document}

\title{Superconductivity near a nematoelastic quantum critical point}

\author{Vanuildo S. de Carvalho}
\affiliation{Instituto de F\'{i}sica, Universidade Federal de Goi\'as, 74.001-970, Goi\^ania-GO, Brazil}
\author{Hermann Freire}
\affiliation{Instituto de F\'{i}sica, Universidade Federal de Goi\'as, 74.001-970, Goi\^ania-GO, Brazil}
\date{\today}

\begin{abstract}

We study the pairing instability of a two-dimensional metallic system induced by Ising-nematic quantum fluctuations in the presence of an unavoidable relevant coupling of the nematic order parameter to the elastic modes (acoustic phonons) of the lattice. We find that this nematoelastic coupling $\lambda_\mathrm{latt}$ leads to a decrease of both the superconducting (SC) critical temperature $T_c$ and the gap function $\Delta$, regardless of the gap symmetry. Interestingly, we show that $\lambda_\mathrm{latt}$ provides a knob that allows us to investigate the emergence of the SC phase at low temperatures, as an instability from either a non-Fermi liquid or a Fermi liquid normal state. The phase transitions between the SC and these normal states are characterized by different critical exponents, which may also vary for each gap symmetry. Finally, we argue that these results might explain the dependence of $T_c$ in the vicinity of the nematic quantum critical point exhibited by the compound FeSe$_{1 - x}$S$_x$.
\end{abstract}

\maketitle

\emph{\textcolor{blue}{Introduction.--}} A theoretically controlled analysis of two-dimensional metals near an Ising-nematic quantum critical point (QCP) remains a big challenge nowadays in the field of strongly correlated systems. It is now well-established that the renormalization group approach with large-$N$ expansion eventually breaks down for this theory at low enough energies \cite{Lee-PRB(2009), Metlitski-PRBa(2010),*Metlitski-PRBb(2010), Senthil-PRB(2010), Lee-ARCMP(2018)}. The Ising-nematic phase refers to a long-range instability with $\mathbf{Q} = 0$ momentum transfer that leads to the lowering of the point-group rotational symmetry, while preserving the properties of the underlying system with respect to translational symmetry \cite{Fradkin-ARCMP(2010),Fernandes-NP(2014)}. Recently, both analytical and numerical approaches \cite{Lee-PRB(2009), Metlitski-PRBa(2010),Senthil-PRB(2010),Hartnoll-PRB(2014),Metlitski-PRB(2015),Lederer-PRL(2015),Berg-PRX(2016),Lederer-PNAS(2017), Lee-ARCMP(2018), Berg-ARCMP(2019)} have shown that the emergence of such a phase could potentially explain some of the physical properties observed in the phase diagram of high-temperature superconductors, such as, e.g., the cuprates \cite{Hinkov-S(2008),Taillefer-N(2010),Matsuda-NP(2017)}, the iron-based superconductors \cite{Chu-S(2010),Kasahara-N(2012),Hosoi-PNAS(2016),Licciardello-N(2019),Coldea-NP(2019)}, and also in other correlated materials \cite{Ronning-NP(2017),Jarillo-Sci(2021)}.

Although the qualitative comparison between the theoretical predictions with experimental data in these systems is reasonably good, a clear picture regarding the impact of electronic nematicity on the corresponding phase diagrams is still lacking. This occurs because for most systems this instability appears along with the emergence of other long-range orders such as antiferromagnetism and superconductivity, which complicate the analysis. In this regard, it was recently found that the iron-based superconductors FeSe$_{1 - x}$S$_x$ \cite{Hosoi-PNAS(2016)} and LaFeAsO$_{1 - x}$F$_x$ \cite{Yang-CPL(2015)} display a quantum phase transition (QPT) to an electronic nematic state disentangled from other magnetic or charge ordered phases, and thus they constitute the most promising platform to study the effects of electronic nematicity. 

However, a more realistic attempt to investigate such a quantum critical theory should also include the unavoidable interaction of the nematic order-parameter with the elastic modes (acoustic phonons) of the lattice \cite{Cowley-PRB(1976),Xu-PRB(2009),Fernandes-PRL(2010),Paul-PRB(2010),Paul-PRB(2017),Schmalian-PRB(2016)}. By taking into account this relevant interaction, previous works have predicted that the low-temperature thermodynamic \cite{Paul-PRL(2017)} and transport \cite{Carvalho-PRB(2019), Wang-PRB(2019), Freire-AP(2020)} properties of the corresponding quantum critical state have a tendency of becoming Fermi-liquid-like below an energy scale set by the nematoelastic interaction. This happens because the shear strain associated with this interaction constrains the quantum critical region to a few high symmetry directions \cite{Paul-PRB(2017)}. Remarkably, the authors of Refs. \cite{Coldea-NP(2019),Bristow-PRR(2020)} have shown that in the vicinity of the nematic QCP, the electrical resistivity clearly transitions from a non-Fermi liquid (NFL) to a Fermi liquid (FL) regime as the temperature is lowered, in good agreement with the theoretical results that consider the effects of the nematoelastic coupling \cite{Carvalho-PRB(2019), Wang-PRB(2019), Freire-AP(2020)}.

\begin{figure}[t]
\centering
\includegraphics[width = 0.97\linewidth]{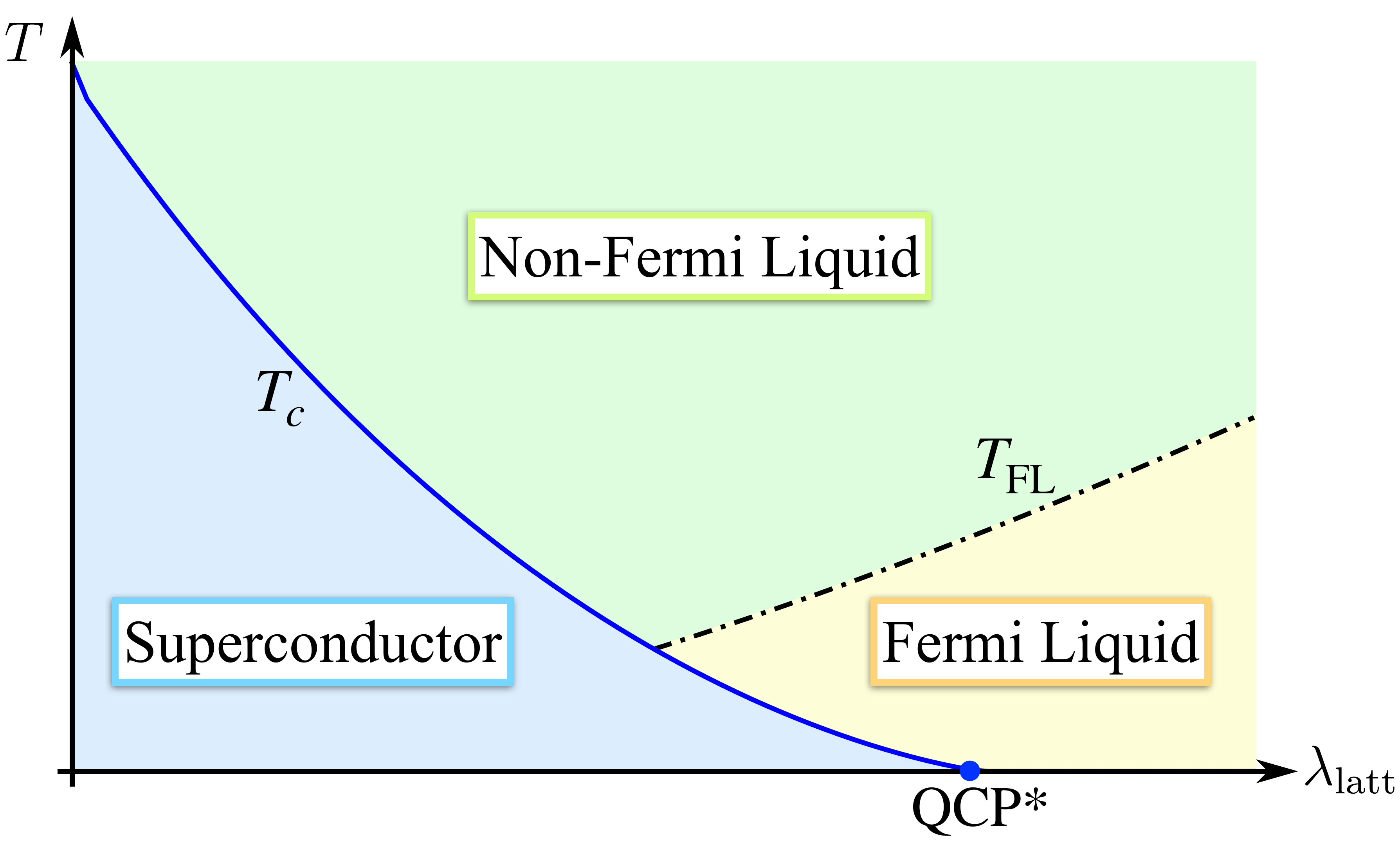}
\caption{Phase diagram of a 2D metallic system at the onset of Ising-nematicity when the nematic degrees of freedom couple to the acoustic phonons by means of an effective nematoelastic interaction $\lambda_\mathrm{latt} \sim g^2_\mathrm{latt}$. In the regime where $\lambda_\mathrm{latt}$ is negligible compared to the effective electron interaction $\gamma \sim g^2$, the SC state emerges out of a non-Fermi liquid (NFL). On the other hand, as $\lambda_\mathrm{latt}$ approaches a lattice-induced quantum critical point (QCP$^*$), whose location depends on $\gamma$ as a power-law, the normal state from which SC emerges is a Fermi liquid (FL). The solid line denoted by $T_c$ refers to the transition to the SC state calculated in the present work [see Eq. \eqref{Eq_SC_Tc}]. The dotted-dashed line represents a crossover between the FL and the NFL regimes, which is given by $T_\mathrm{FL} \sim \lambda^{3/2}_\mathrm{latt} E_F$ \cite{Paul-PRL(2017)}, where $E_F$ is the Fermi energy.}\label{Phase_Diagram}
\end{figure}

In this Letter, we investigate the impact of the nematoelastic coupling, $g_\mathrm{latt}$, on a two-dimensional (2D) SC state that emerges near the onset of an Ising-nematic order. In order to do that, we apply the strong-coupling Eliashberg theory to obtain the critical SC temperature $T_c$ and the gap function $\Delta(\theta, \omega_n)$ around a circular Fermi surface at $T = T_c$, by considering both \emph{s}-wave and \emph{d}-wave symmetries. In this connection, we highlight two recent works that address similar issues regarding this problem, which can be viewed as complementary to our present study (see Refs. \cite{Paul-PRB(2017), Klein-PRB(2018), *Klein-NPJ(2019)}). The first one \cite{Paul-PRB(2017)} describes, within a weak-coupling BCS theory, how SC emerges near a nematic QCP for a system in the presence of such a nematoelastic interaction. The authors find that $T_c$ only increases if the nematoelastic coupling is strong enough and the system is dominated by non-nematic interactions \cite{Paul-PRB(2017)}. By contrast, the second work \cite{Klein-PRB(2018)} focus on the properties of the SC state driven by strong nematic fluctuations (but without the nematoelastic coupling) from a Eliashberg-theory perspective. As a result, they find that nematic fluctuations always favor an \emph{s}-wave SC state, despite it becoming almost degenerate with a \emph{d}-wave one in the weak-coupling regime \cite{Klein-PRB(2018)}.

Our main results are summarized in the phase diagram of Fig. \ref{Phase_Diagram}. The increase of $\lambda_\mathrm{latt} \sim g^2_\mathrm{latt}$ is always detrimental to superconductivity and, therefore, reduces $T_c$ and the $\Delta(\theta, \omega_n)$. This behavior is qualitatively similar for both \emph{s}-wave and \emph{d}-wave pairing, although the reduction in $T_c$ and $\Delta(\theta, \omega_n)$ becomes stronger in the latter case. In the limit where $\lambda_\mathrm{latt}$ is in the vicinity of a lattice-induced quantum critical point (QCP$^*$), whose location depends on the effective electron interaction $\gamma \sim g^2$ as a power-law, the system undergoes a SC-FL quantum phase transition. In addition, the effective coupling $\lambda_\mathrm{latt}$ allows us to obtain SC states emerging from different metallic states, represented by the FL and NFL phases. This leads to interesting new predictions, because the NFL-SC and the FL-SC phase transitions turn out to be described in terms of different critical exponents, which may also vary for each gap symmetry. As a result, we argue that the suppression of $T_c$ in the vicinity of the nematic QCP for the compound FeSe$_{1 - x}$S$_x$ \cite{Coldea-NP(2019)} may be interpreted as a fingerprint of nematoelastic quantum criticality.

\emph{\textcolor{blue}{Model.--}} We model the instability to the nematic state in a 2D system in terms of the coupling of the nematic order parameter to the electronic quasiparticles \cite{Lee-PRB(2009), Metlitski-PRBa(2010)}. We include the influence of the lattice on that order parameter through a linear coupling to the local orthorhombic strain $\epsilon(\mathbf{r}) \equiv \epsilon_{xx}(\mathbf{r}) - \epsilon_{yy}(\mathbf{r})$  \cite{Xu-PRB(2009),Fernandes-PRL(2010),Paul-PRB(2010),Paul-PRB(2017),Schmalian-PRB(2016)}, where the strain is defined in terms of the displacement vector $\bm{u}(\mathbf{r})$ by means of $\epsilon_{ij} = \partial_{i}u_{j} + \partial_{j}u_{i}$ \cite{Landau-PP(1970),Cowley-PRB(1976)}. The minimal Hamiltonian describing this system reads
\begin{align}
H & = H_\text{n-elec} + H_\text{n-latt}, \label{Eq_Ham_01}\\
H_\text{n-elec} & = \sum\limits_{\mathbf{k}, \sigma} \xi_\mathbf{k} \psi^{\dagger}_\sigma(\mathbf{k}) \psi_\sigma(\mathbf{k}) + \sum_{\mathbf{q}} \Omega(\mathbf{q}) \varphi(\mathbf{q}) \varphi(- \mathbf{q}) \nonumber \\
& + g \sum\limits_{\mathbf{q}, \mathbf{k}, \sigma} f(\mathbf{k}) \psi^{\dagger}_\sigma\left(\mathbf{k} - \frac{\mathbf{q}}{2} \right) \psi_\sigma\left(\mathbf{k} + \frac{\mathbf{q}}{2} \right) \varphi(\mathbf{q}), \label{Eq_Ham_02}\\ 
H_\text{n-latt} & = \sum\limits_{\mathbf{q} \neq 0} \left[ \frac{1}{2}\bm{u}^\dagger_\mathbf{q} \boldsymbol{\mathcal{N}}(\mathbf{q}) \bm{u}_\mathbf{q} + i g_\mathrm{latt}  \mathbf{a}_\mathbf{q} \cdot \bm{u}_\mathbf{q} \varphi(- \mathbf{q}) \right], \label{Eq_Ham_03}
\end{align}
where $\psi^{\dagger}_\sigma(\mathbf{k})$ ($\psi_\sigma(\mathbf{k})$) is the creation (annihilation) operator for fermions with spin projection $\sigma \in \{ \uparrow, \downarrow \}$ and band dispersion $\xi_\mathbf{k}$, $\Omega_{\mathbf{q}}$ is the dispersion of the nematic bosons described by $\varphi(\mathbf{q})$, $g$ and $g_\mathrm{latt}$ are, respectively, the nematic and the nematoelastic interactions, and $f(\mathbf{k}) \equiv (\cos k_x - \cos k_y)$ denotes a \emph{d}-wave form factor for a nematic state with $B_{1g}$ symmetry. Moreover, the acoustic phonons are described in terms of the Fourier-transformed displacement vector $\bm{u}_\mathbf{q}$, $\mathbf{a}_{\mathbf{q}} = (q_{x}, - q_{y}, 0)$ is a 2D vector, and $\boldsymbol{\mathcal{N}}(\mathbf{q})$ stands for the matrix of the elastic constants $C_{ij}$ \cite{Landau-PP(1970),Cowley-PRB(1976)}, which for a 2D system with tetragonal symmetry is given by
\begin{equation}
\boldsymbol{\mathcal{N}}(\mathbf{q}) = \begin{pmatrix}C_{11}q_{x}^{2} + C_{66}q_{y}^{2} & (C_{12} + C_{66})q_{x}q_{y} \\
(C_{12} + C_{66})q_{x}q_{y} & C_{66}q_{x}^{2} + C_{11}q_{y}^{2}
\end{pmatrix}.
\end{equation}

\begin{figure*}[t]
\centering
\centering \includegraphics[width=0.325\linewidth,valign=t]{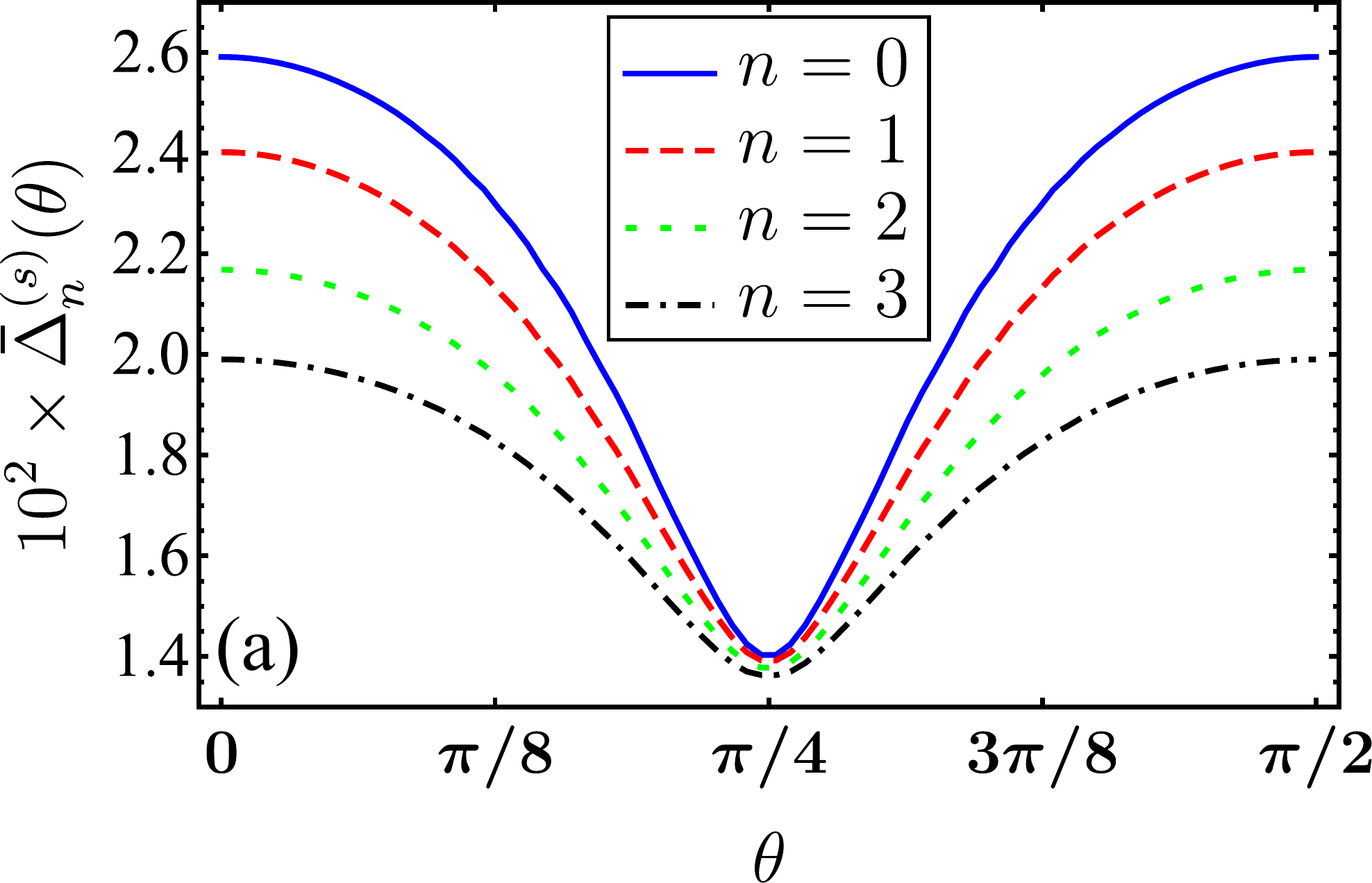} \hfil{} \includegraphics[width=0.325\linewidth,valign=t]{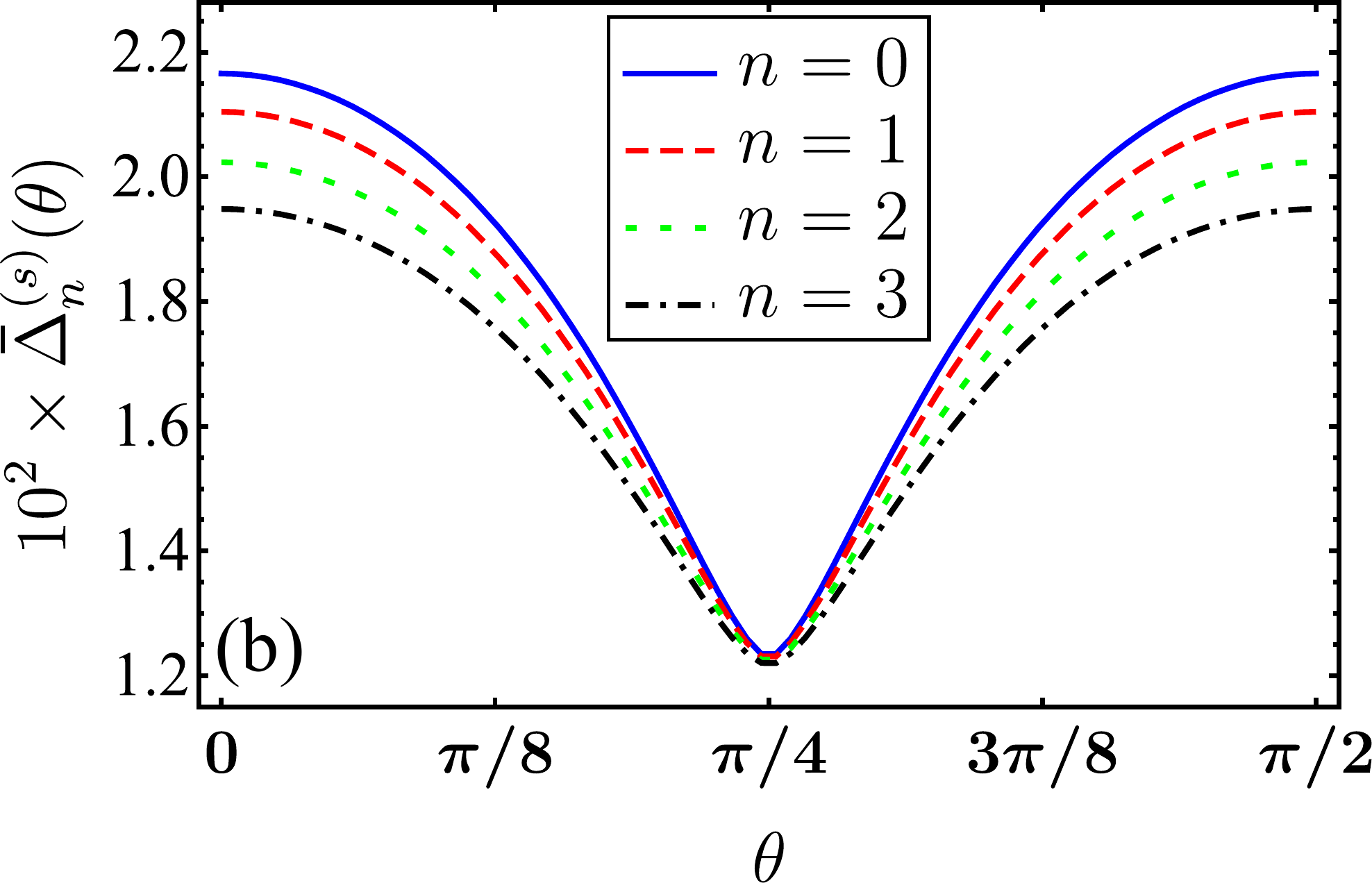} \hfil{} \includegraphics[width=0.325\linewidth,valign=t]{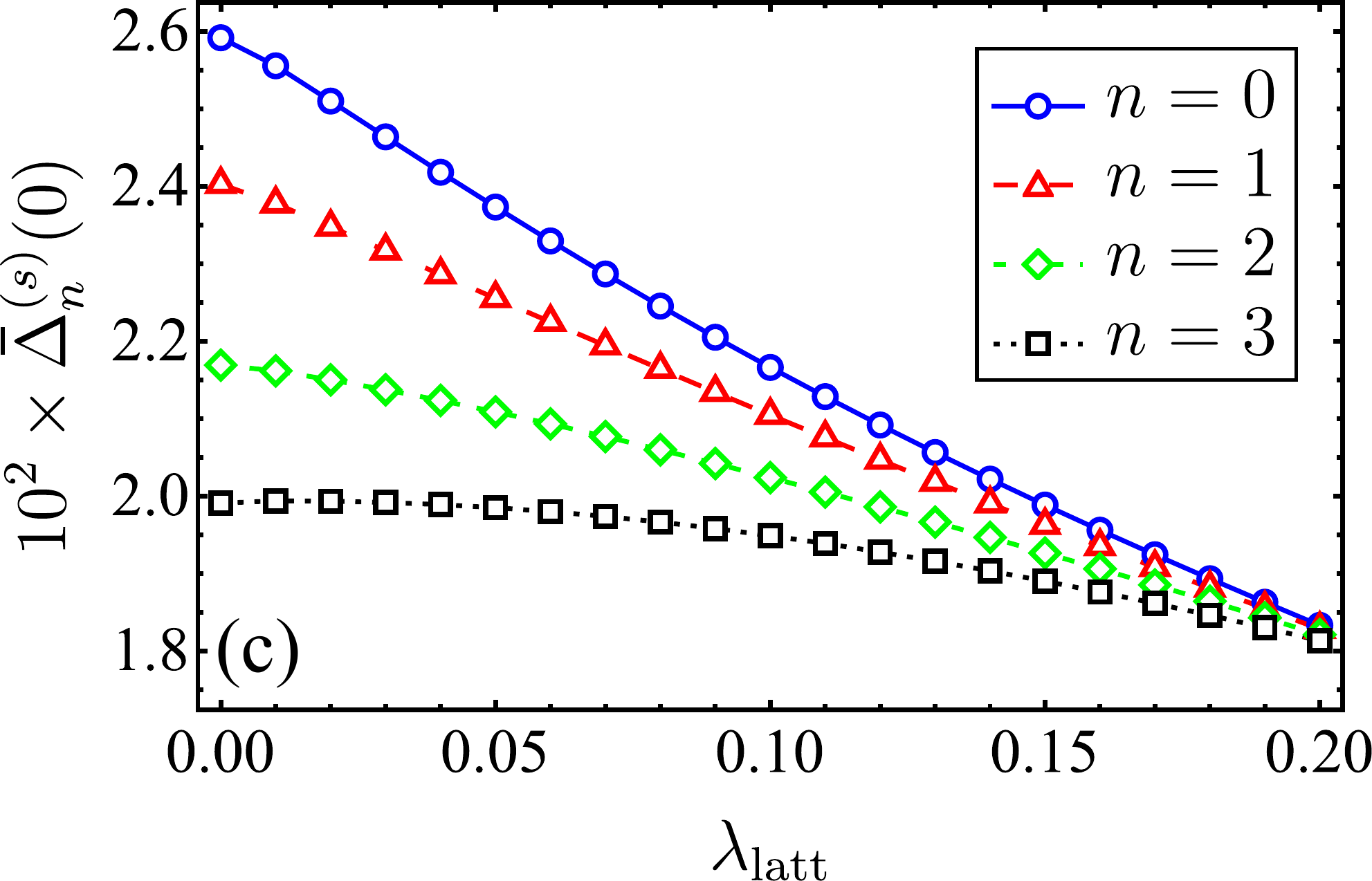} \vfil{} \includegraphics[width=0.325\linewidth,valign=t]{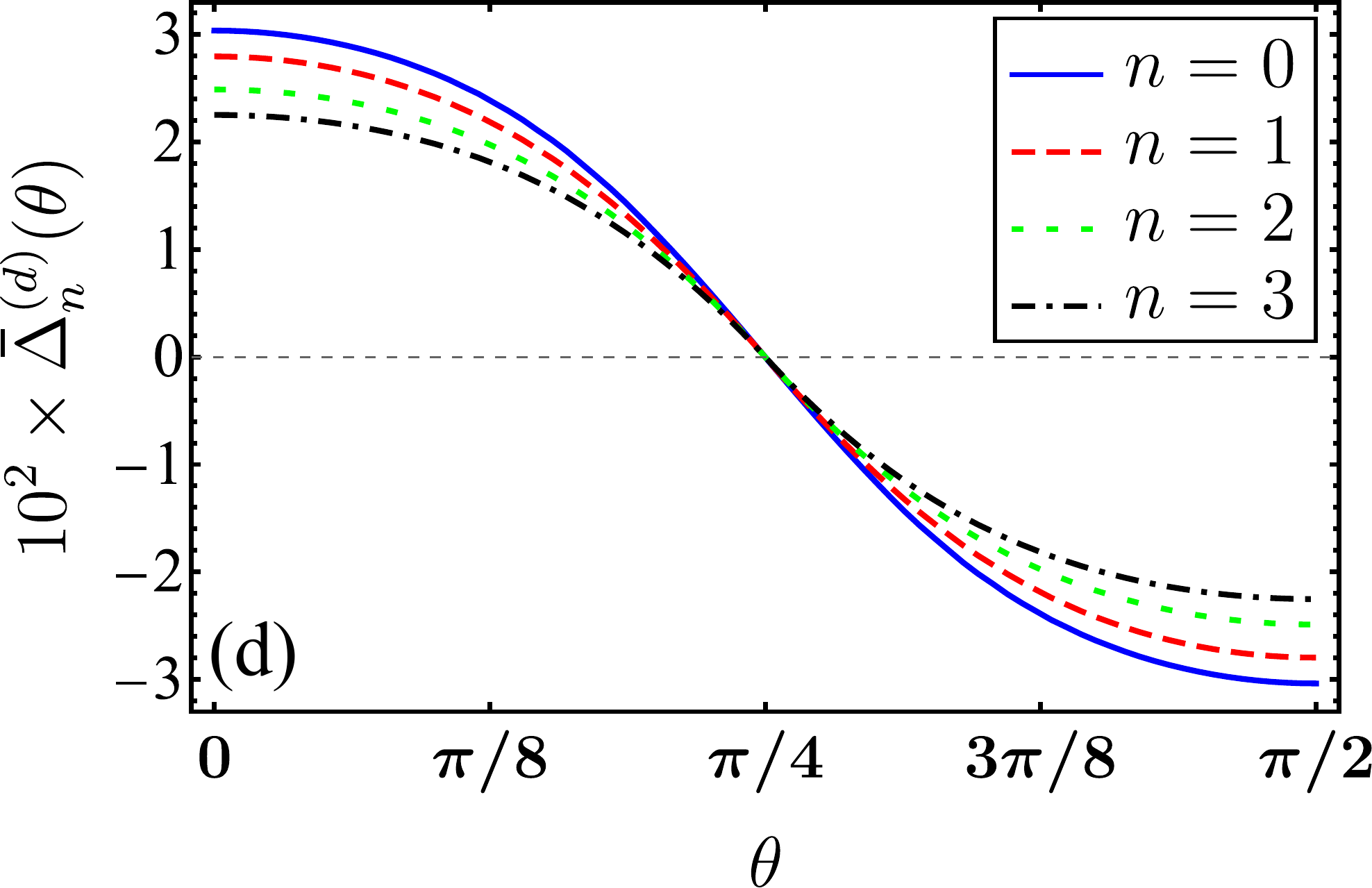} \hfil{} \includegraphics[width=0.325\linewidth,valign=t]{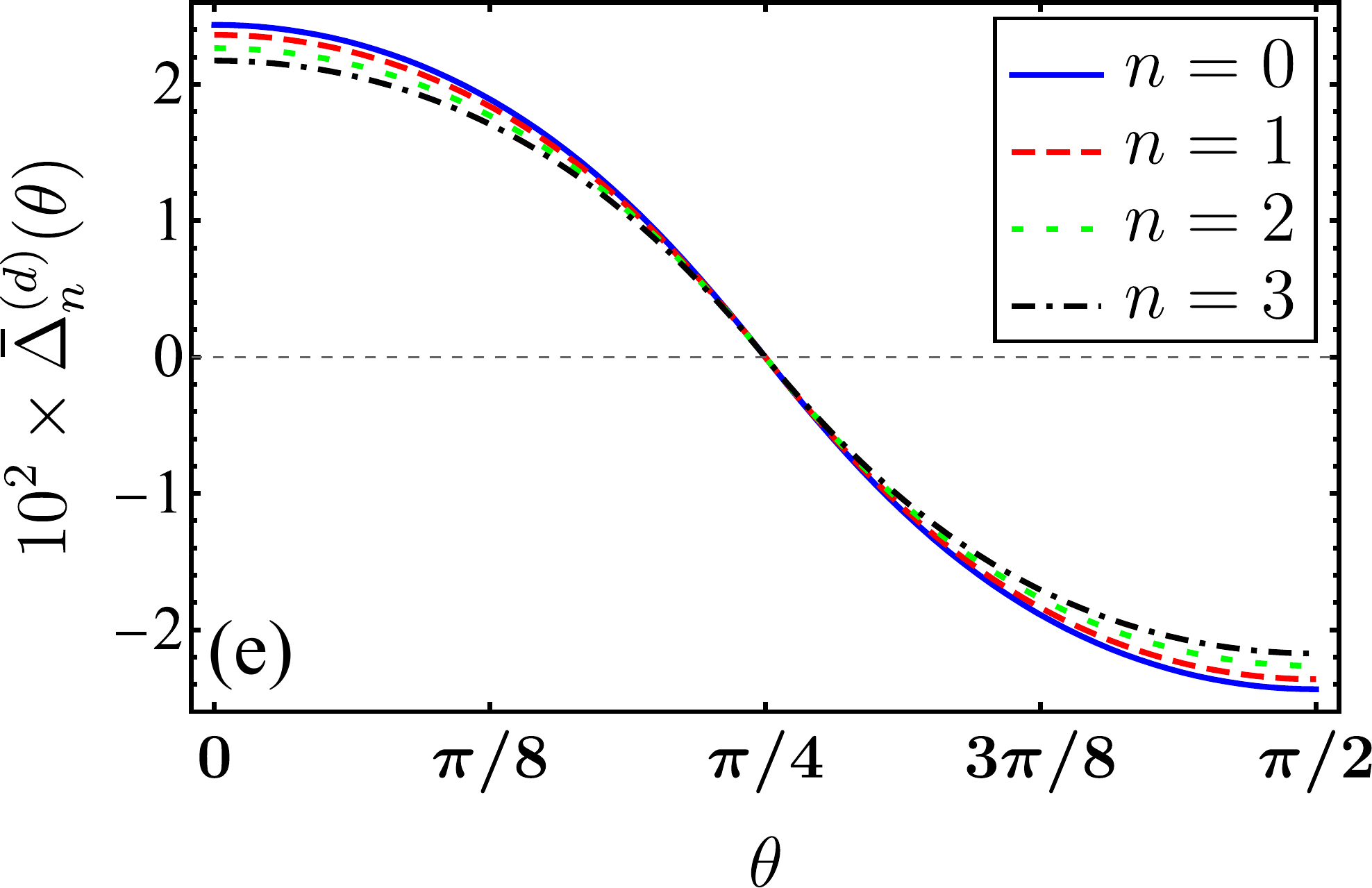} \hfil{} \includegraphics[width=0.325\linewidth,valign=t]{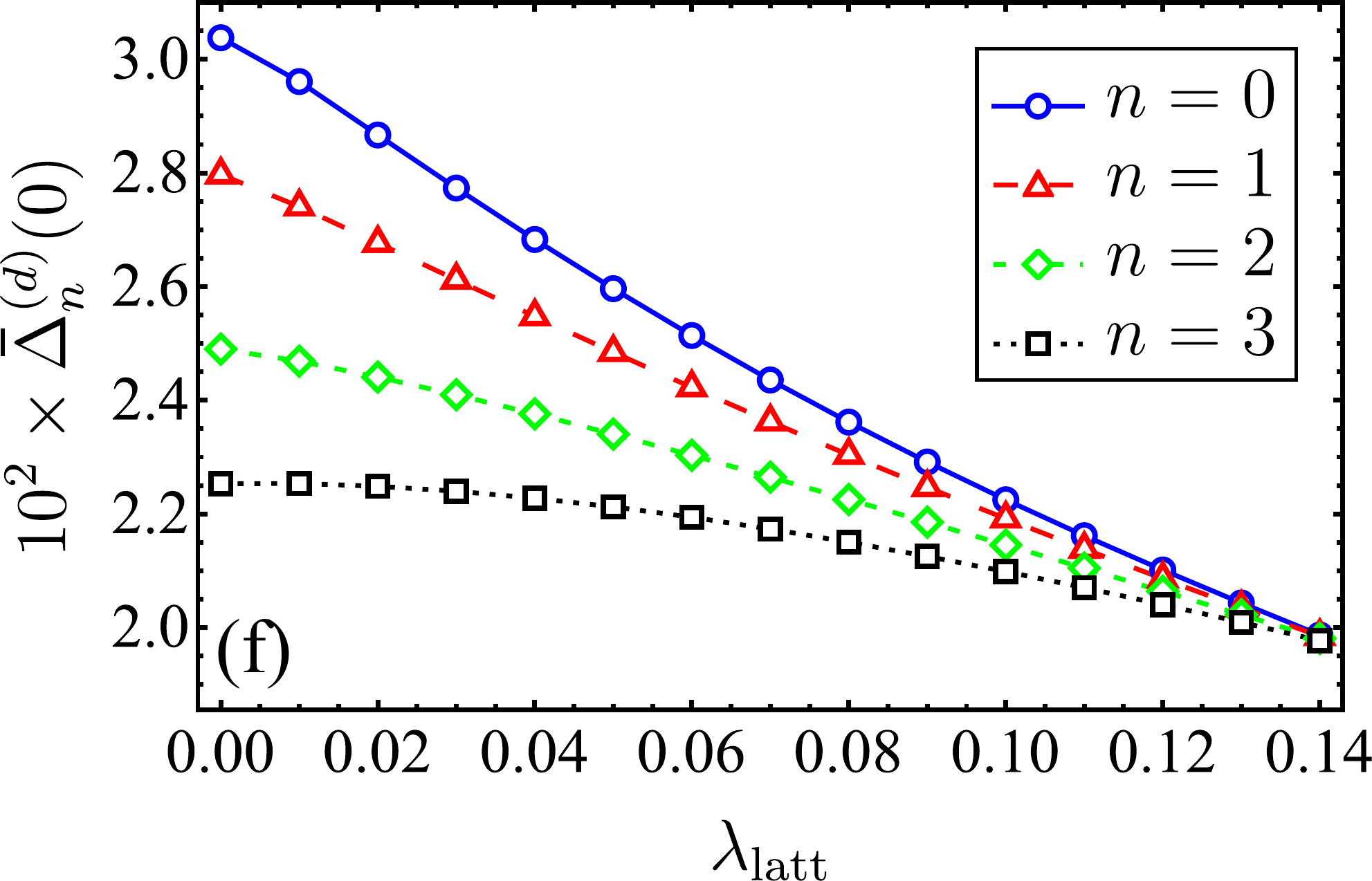} 
\caption{(a)-(b) Behavior of the SC gap (in units of the Fermi energy) with \emph{s}-wave symmetry along the Fermi surface and for Matsubara frequencies $\omega_n = (2n +1) \pi T$ obtained by setting $\gamma = 0.40$ and then solving the linearized Eliashberg equations at $T = T_c$ for (a) $\lambda_\mathrm{latt} = 0$ and (b) $\lambda_\mathrm{latt} = 0.10$. Notice that the effect of increasing $\lambda_\mathrm{latt}$ is to diminish $\bar{\Delta}^{(s)}_n(\theta)$ along the whole Fermi surface. (c) The SC gap at the hot spot $\theta = 0$ decreases monotonically with $\lambda_\mathrm{latt}$. At $\lambda_\mathrm{latt} = \lambda_{\mathrm{latt}, c}$, which is given here by $\lambda_{\mathrm{latt}, c} \approx 0.21$, the gap $\bar{\Delta}^{(s)}_n(\theta = 0)$ becomes degenerate in both frequency and momentum. Panels (d)-(f) show the behavior of a SC gap with \emph{d}-wave symmetry which arises for $\gamma = 0.40$ and then fixing (d) $\lambda_\mathrm{latt} = 0$, and (e) $\lambda_\mathrm{latt} = 0.18$. (f) Behavior of $\bar{\Delta}^{(d)}_n(\theta)$ for $\theta = 0$ as a function of $\lambda_\mathrm{latt}$ and $\omega_n$. Notice that the \emph{s}-wave and \emph{d}-wave SC gap display similar features as $\lambda_\mathrm{latt}$ approaches $\lambda_{\mathrm{latt}, c}$.}\label{Gap_Dependence}
\end{figure*}

In order to obtain the effective field description of the nematic interaction between the itinerant fermions, we consider the one-loop correction of the nematic propagator by the particle-hole polarization bubble, in addition to integrating out the acoustic phonons from the action corresponding to the Hamiltonian $H_\text{n-latt}$. As a result, if we set the density of states to its Fermi level value $N_F$, the bosonic propagator at the bare QCP distance $r_0$ evaluates to $D^{-1} = N^{-1}_F ( r_0 + K q^2 + \Pi_\mathrm{elec} ) + \Pi_\mathrm{latt}$, where $\Pi_\mathrm{elec}$ and $\Pi_\mathrm{latt}$ refer, respectively, to the particle-hole and lattice polarization bubbles. They are given by
\begin{align}
\Pi_\mathrm{elec}(\mathbf{q}, \Omega_m) & = (N_F g)^2 \cos^2(2 \varphi) \frac{|\Omega_m|}{v_F q}, \\
\Pi_\mathrm{latt}(\mathbf{q}, \Omega_m) & =  \frac{(\gamma_2 + \gamma_3)g^2_\mathrm{latt}}{\gamma^2_1 - \Upsilon^2(\varphi)} \hspace{-0.1cm} \bigg[ \cos^2(2 \varphi) - \frac{\gamma_1 + \gamma_3}{\gamma_2 + \gamma_3} \bigg].
\end{align}
Furthermore, the parameters in the above equation turn out to be $\gamma_{1} = C_{11} + C_{66}$, $\gamma_{2} = C_{11} - C_{66}$, $\gamma_{3} = C_{12} + C_{66}$, and $\Upsilon\left(\varphi\right)=\sqrt{\frac{1}{2}(\gamma_{2}^{2}+\gamma_{3}^{2})+\frac{1}{2}(\gamma_{2}^{2}-\gamma_{3}^{2})\cos(4\varphi)}$ \cite{Carvalho-PRB(2019)}. In what follows, we consider systems in which the lattice is equally strong to distortion fluctuations in both the $B_{1g}$ and $B_{2g}$ channels. This implies that $(C_{11} - C_{12})/2 = C_{66}$ and, consequently, leads to $\gamma_2 = \gamma_3$. Therefore, the bosonic nematic mass becomes angular dependent and given by $r(\varphi) = (r_0 - r_{0, c})/(K k^2_F) + \lambda_\mathrm{latt} \cos^2(2 \varphi)$, where $r_{0, c} \equiv \frac{N_F g_{\mathrm{latt}}^{2}}{C_{66}}$ and $\lambda_\mathrm{latt} \equiv \frac{r_{0, c}}{2 K k^2_F} \left(1 + \frac{C_{12}}{C_{11}} \right)$. Since $r(\varphi)$ goes to zero only along the diagonals of the Brillouin zone, the phenomenon of nematic quantum criticality becomes directional-selective (see Ref. \cite{Paul-PRL(2017)}).

\emph{\textcolor{blue}{Self-consistent Eliashberg equations.--}} The emergence of the SC state in the present model is studied in terms of the Migdal-Eliashberg theory \cite{Eliashberg-JETP(1960), Scalapino-PR(1966), Carbotte-RMP(1990)}. As expressed in Eqs. \eqref{Eq_Ham_01}--\eqref{Eq_Ham_03}, we consider that the attractive interaction between the conduction electrons is provided solely by the collective fluctuations of the nematic field $\varphi(\mathbf{q})$, although we note that other effects such as the Coulomb repulsion and the electron-phonon interaction might also become relevant in the limit in which lattice distortions are strong. In order to derive the Eliashberg equations, we first express the fermionic operators $\psi_\sigma(\mathbf{k})$ and $\psi^\dagger_\sigma(\mathbf{k})$ in terms of the Nambu spinor $\Psi(\mathbf{k}) = ( \psi_{\uparrow}(\mathbf{k}), \; \psi^{\dagger}_{\downarrow}(-\mathbf{k}) )^T$. As a result, the interacting fermionic propagator related to $\Psi(\mathbf{k})$ becomes $\boldsymbol{\mathcal{G}}^{-1}(\mathbf{k}, \omega_n) = i\omega_n \boldsymbol{\tau}^0 - \xi_\mathbf{k}\boldsymbol{\tau}^z - \boldsymbol{\Sigma}(\mathbf{k}, \omega_n)$, where $\boldsymbol{\tau}^0$ and $\boldsymbol{\tau}^{x, y, z}$  are, respectively, the identity and Pauli matrices in the Nambu space, and $\boldsymbol{\Sigma}(\mathbf{k}, \omega_n)$ refers to the self-energy of the conduction electrons. Taking into account the effect of vertex interactions with bosonic propagators renormalized by one-loop particle-hole and lattice-distortion corrections, the fermionic self-energy evaluates to
\begin{align}\label{Eq_Self_Energy_01}
\boldsymbol{\Sigma}(\mathbf{k}, \omega_n) & = g^2 T \sum_{\mathbf{k}', n'} f^2 \bigg(\frac{\mathbf{k} + \mathbf{k}'}{2} \bigg) D(\mathbf{k} - \mathbf{k}', \omega_n - \omega_{n'}) \nonumber \\
& \times \boldsymbol{\tau}^z \boldsymbol{\mathcal{G}}(\mathbf{k}', \omega_{n'}) \boldsymbol{\tau}^z.
\end{align}
The above result is formally obtained by summing an infinite series involving rainbow Feynman diagrams with non-interacting fermionic propagators. 

The parametrization of the fermionic the self-energy in Eq. \eqref{Eq_Self_Energy_01} according to $\boldsymbol{\Sigma}(\mathbf{k}, \omega_n) = i\omega_n[1 - Z_n(\mathbf{k})]\boldsymbol{\tau}^0 + \chi_n(\mathbf{k})\boldsymbol{\tau}^z + \Phi_n(\mathbf{k})\boldsymbol{\tau}^x$ yields three coupled Eliashberg equations relating the mass renormalization function $Z_n(\mathbf{k})$, the energy shift $\chi_n(\mathbf{k})$, and the anomalous self-energy $\Phi_n(\mathbf{k})$ that encodes the information on the pairing mechanism. To simplify our analysis, we restrict the momentum dependence of $Z_n(\mathbf{k})$, $\chi_n(\mathbf{k})$, and $\Phi_n(\mathbf{k})$ to the angle $\theta$ of the Fermi momentum $\mathbf{k}_\theta = k_F(\cos\theta \hat{\mathbf{x}} + \sin\theta \hat{\mathbf{y}})$ and evaluate the momentum integrals perpendicular to the Fermi surface by setting the density of states to $N_F$, according to the Migdal-Eliashberg approximation. This assumption is reasonable because the nematic boson is assumed to be a slow mode compared to a fermion, i.e., the boson dispersion is much smaller than its fermionic counterpart. We also make the linearization of these equations with respect to the anomalous self-energy $\Phi(\theta)$, because this approximation is sufficient to determine the $T_c$ and the gap dependence on $\theta$ at the SC transition. As a result, we obtain that the energy shift becomes identically zero, i.e., $\chi_n(\theta) = 0$. The remaining linearized Eliashberg equations evaluate to
\begin{align}
Z_n(\theta) - 1 = & \frac{\pi \overline{g}^2}{2 N_F} \frac{T}{\omega_n} \sum\limits_{\omega_{n'}} \int^{2\pi}_{0} \frac{d\theta'}{2\pi} D(\mathbf{k}_{\theta} - \mathbf{k}'_{\theta'}, \omega_n - \omega_{n'}) \nonumber \\
& \times f^2\bigg(\frac{\mathbf{k}_{\theta} + \mathbf{k}'_{\theta'}}{2} \bigg) \operatorname{sgn}(\omega_{n'}), \label{Eq_Linear_Eliashberg_01}\\
Z_n(\theta)\Delta_n(\theta) = & \frac{\pi \overline{g}^2}{2 N_F} T \sum\limits_{\omega_{n'}} \int^{2\pi}_{0} \frac{d\theta'}{2\pi} D(\mathbf{k}_{\theta} - \mathbf{k}'_{\theta'}, \omega_n - \omega_{n'}) \nonumber \\
& \times f^2\bigg(\frac{\mathbf{k}_{\theta} + \mathbf{k}'_{\theta'}}{2} \bigg) \frac{\Delta_{n'}(\theta')}{|\omega_{n'}|}, \label{Eq_Linear_Eliashberg_02}
\end{align}
where $\overline{g} \equiv N_F g$ and $\Delta_n(\theta) \equiv \Phi_n(\theta)/Z_n(\theta)$ corresponds to the SC gap. Therefore, by inserting Eq. \eqref{Eq_Linear_Eliashberg_01} into \eqref{Eq_Linear_Eliashberg_02}, this reduces the pairing problem to the solution of a 2D integral equation for $\Delta_n(\theta)$.

\begin{figure*}[t]
\centering
\centering \includegraphics[width=0.33\linewidth,valign=t]{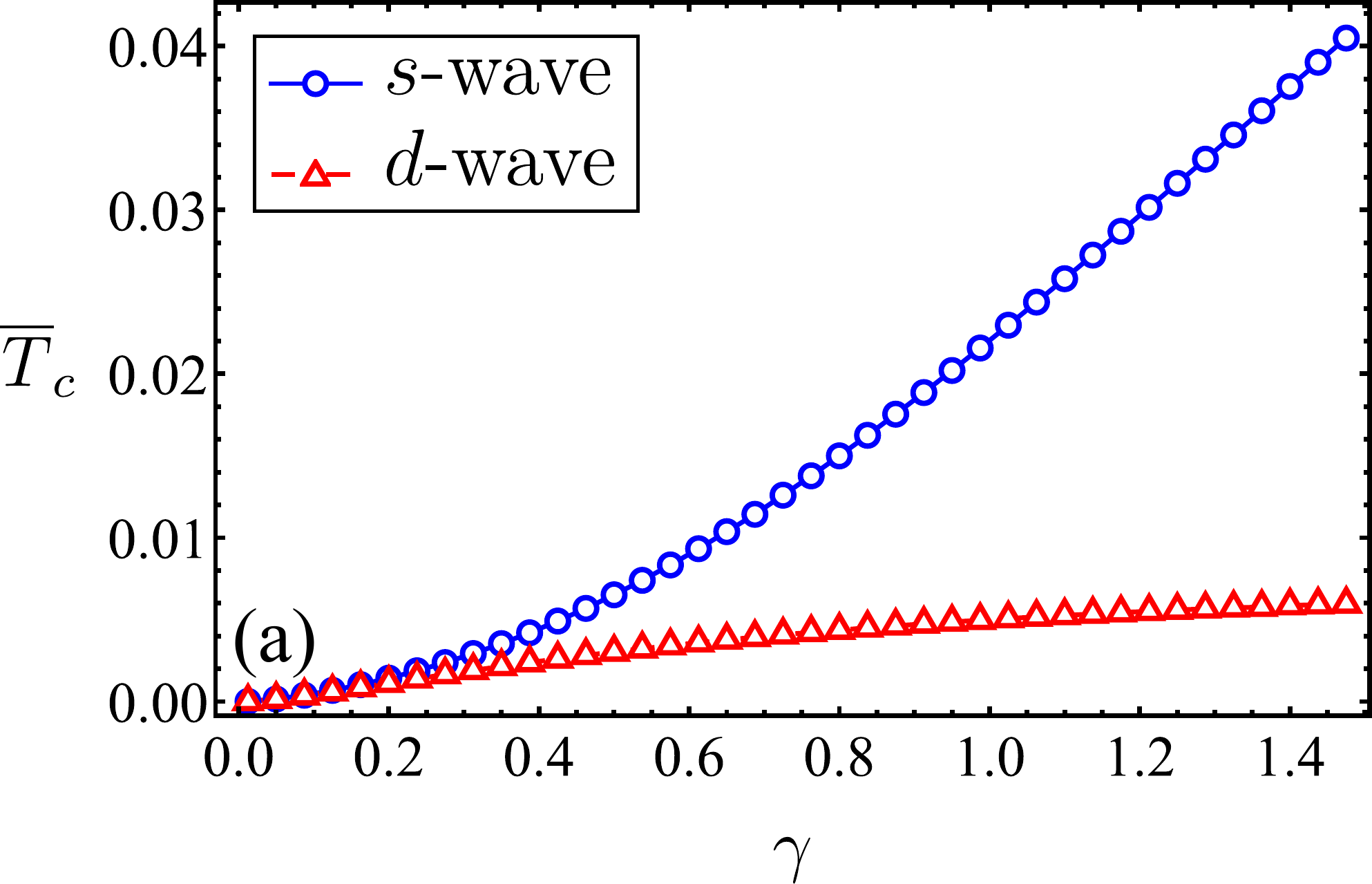} \hfil{} \includegraphics[width=0.31\linewidth,valign=t]{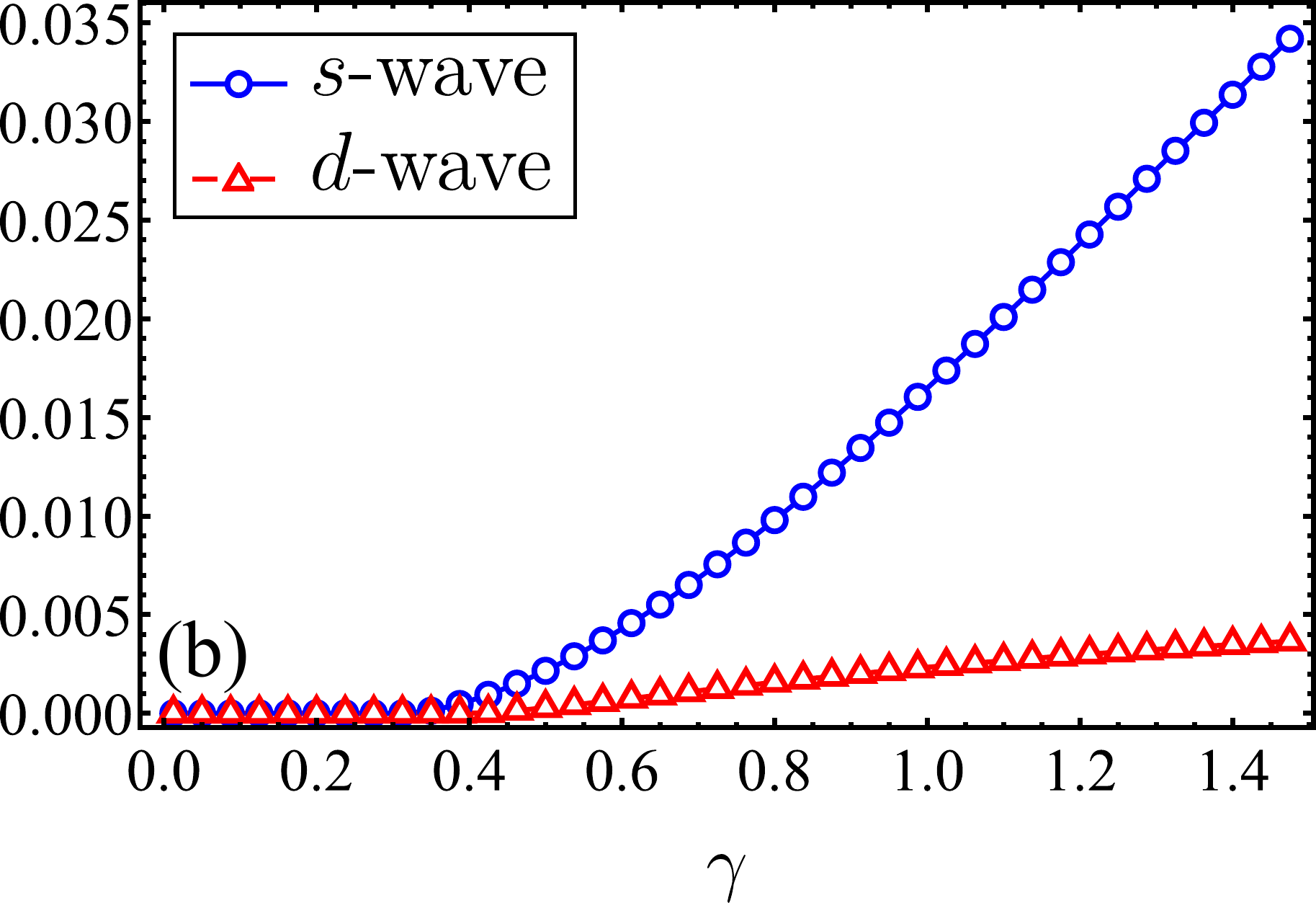} \hfil{} \includegraphics[width=0.335\linewidth,valign=t]{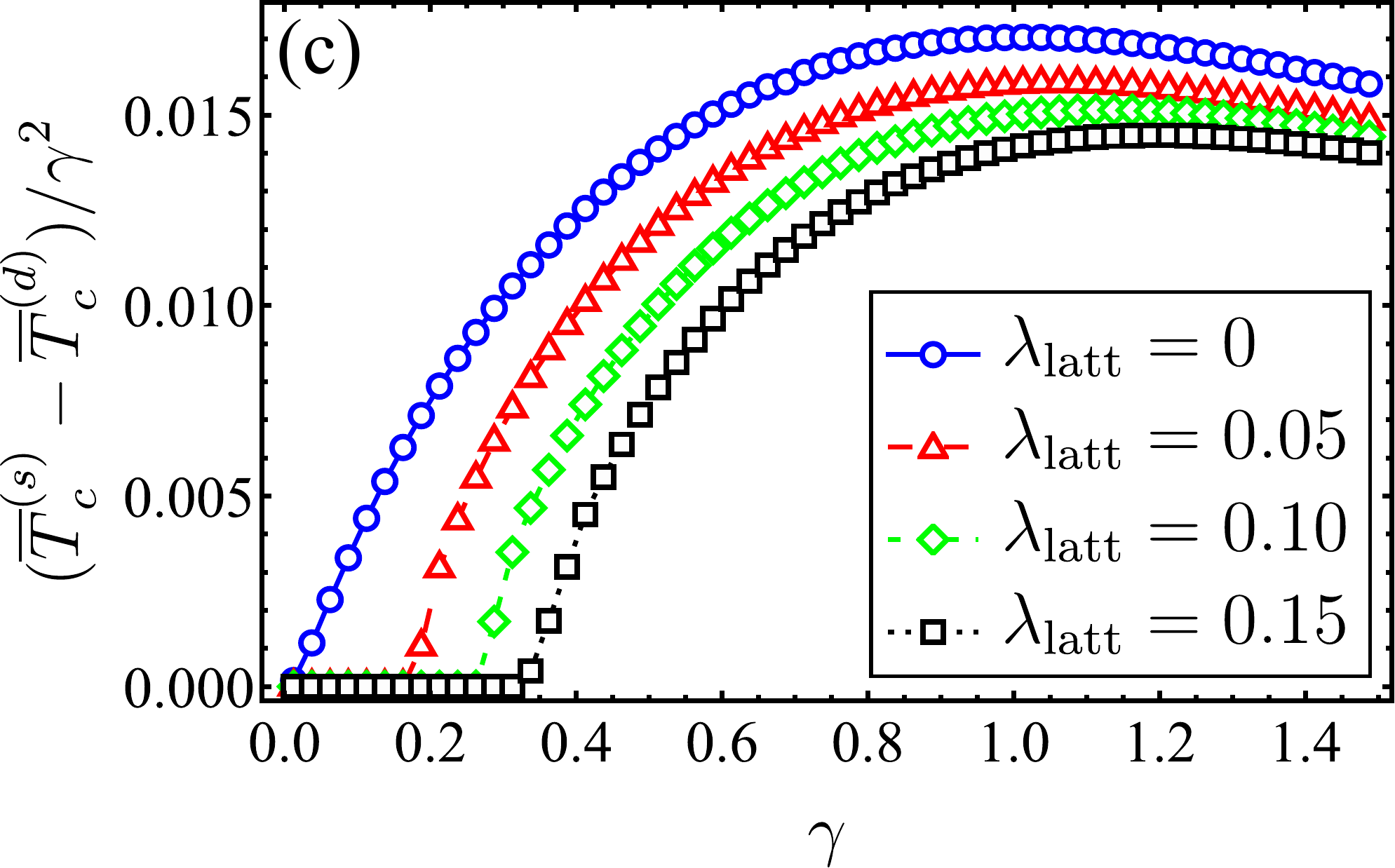}
\caption{SC critical temperature $\overline{T}_c \equiv T_c/E_F$ as a function of the effective electron interaction represented by $\gamma$, and the nematoelastic coupling $\lambda_\mathrm{latt}$. Panels (a) and (b) show the \emph{s}-wave and \emph{d}-wave behavior of $T_c$ for $\lambda_\mathrm{latt} = 0$ and $\lambda_\mathrm{latt} = 0.15$, respectively. Notice that for $\lambda_\mathrm{latt} > 0$, the SC state appears only for $\gamma > \gamma_c$, where the critical interaction $\gamma_c$ increases with $\lambda_\mathrm{latt}$. Besides, the \emph{s}-wave $T_c$ is always larger than its \emph{d}-wave counterpart and has a larger $\gamma_c$ for the same value of the nematoelastic coupling. (c) In the $\gamma \gg \lambda_\mathrm{latt}$ limit, the difference between the \emph{s}-wave and \emph{d}-wave SC critical temperature $T_c$ approaches a constant value, which is independent of $\lambda_\mathrm{latt}$.}\label{Tc_Dependence}
\end{figure*}

We point out here that, instead of approximating the renormalization mass function $Z_n(\theta)$ by its scaling form at zero temperature to solve the gap equation \eqref{Eq_Linear_Eliashberg_02}, we will employ its exact expression at the transition temperature $T_c$. However, this will require the numerical evaluation of the angular integral in Eq. \eqref{Eq_Linear_Eliashberg_01}. On the other hand, the Matsubara sum on the right-hand side of this equation can be put in closed-form. By evaluating it, we obtain
\begin{align}\label{Eq_Linear_Eliashberg_03}
& Z_n(\theta) = 1 - \frac{1}{\overline{\omega}_n}\int_{\theta'}  \bigg\lbrace q(\theta, \theta') \bigg[\psi_0 \bigg(\frac{q(\theta, \theta') \mathcal{C}(\theta, \theta')}{2 \pi \gamma \overline{T} \cos^2(\theta + \theta')} + 1 \bigg) \nonumber \\
& - \psi_0 \bigg(\frac{q(\theta, \theta') \mathcal{C}(\theta, \theta')}{2 \pi \gamma \overline{T} \cos^2(\theta + \theta')} + n + 1\bigg)\bigg] - \frac{\pi \gamma \overline{T} \cos^2(\theta + \theta')}{\mathcal{C}(\theta, \theta')}\bigg\rbrace, 
\end{align}
where $\psi_0(x)$ is the digamma function, $q(\theta, \theta') \equiv 2 |\sin[(\theta - \theta')/2] |$, $\mathcal{C}(\theta, \theta') \equiv r + q^2(\theta, \theta')$, $\gamma = \overline{g}^2/(K k^2_F)$ is the effective electron coupling, $E_F = v_F k_F$ is the Fermi energy, $\overline{T} = T/E_F$ ($\overline{\omega}_n = \omega_n/E_F$) is the reduced temperature (frequency), and $\int_{\theta} (\cdots) \equiv \int^{2 \pi}_{0} \frac{d \theta}{2 \pi} (\cdots)$. Note that to obtain the expression in Eq. \eqref{Eq_Linear_Eliashberg_03}, we made the substitution $f^2[(\mathbf{k}_{\theta} + \mathbf{k}'_{\theta'})/2] = \cos^2(\theta + \theta')$. We should also point out that the distance to the nematic QCP $r = r(\varphi)$ depends on the scattering angles $\theta$ and $\theta'$ through $\varphi = (\theta + \theta')/2 - \pi/2$. Finally, the last term on the right-hand side of Eq. \eqref{Eq_Linear_Eliashberg_03} diverges as the QCP is approached. However, this divergent term is required to cancel out another divergent contribution, which appears in the Matsubara sum of Eq. \eqref{Eq_Linear_Eliashberg_02} when $n' = n$.  

\emph{\textcolor{blue}{Numerical results.--}} The critical temperature $T_c$ and the gap function $\Delta_n(\theta)$ are obtained by solving numerically the integral equation \eqref{Eq_Linear_Eliashberg_02} by the Nystr\"om method \cite{Press-NR(1993)}. In this case, we transform this equation into the eigenvalue problem $\sum_{j, m} \mathcal{K}_{n, m}(\theta_i, \theta_j) \Delta_{m}(\theta_j) = \varepsilon \Delta_{n}(\theta_i)$ and demand that the highest eigenvalue $\varepsilon$ should be equal to zero. All results presented here are obtained by utilizing a cutoff with $N = 30$ Matsubara frequencies and then discretizing the interval $[0, 2 \pi[$ for the variable $\theta$ into 1000 points. For all results appearing in this work, we also consider that the system is located at the QCP emerging from lattice effects, i.e., at $r_0 = r_{0, c}$. Fluctuations of the latter are taken by changing only the effective coupling $\lambda_\mathrm{latt}$. If we set the microscopic interaction $g_\mathrm{latt}$ to a constant, this situation is obtained by fixing the elastic constant $C_{66}$ and then varying either $C_{11}$ or $C_{12}$. 

Figure \ref{Gap_Dependence} describes the behavior of the SC gap for \emph{s}-wave and \emph{d}-wave symmetries at the transition to the SC state, as a function of $\lambda_\mathrm{latt}$. As the nematoelastic interaction is switched on, the SC gap for both symmetries begins to decay along the whole Fermi surface. This implies a monotonic behavior in frequency for the SC gap $\Delta^{(s, d)}_n(\theta)$ at any value of momentum defined by $\theta$. At the critical point $\lambda_\mathrm{latt} = \lambda_{\mathrm{latt}, c}$, where the SC state undergoes a transition to a FL (QCP$^*$), the gap $\Delta^{(s, d)}_n(\theta)$ becomes degenerate in frequency.

\begin{figure*}[t]
\centering
\centering \includegraphics[width=0.48\linewidth,valign=t]{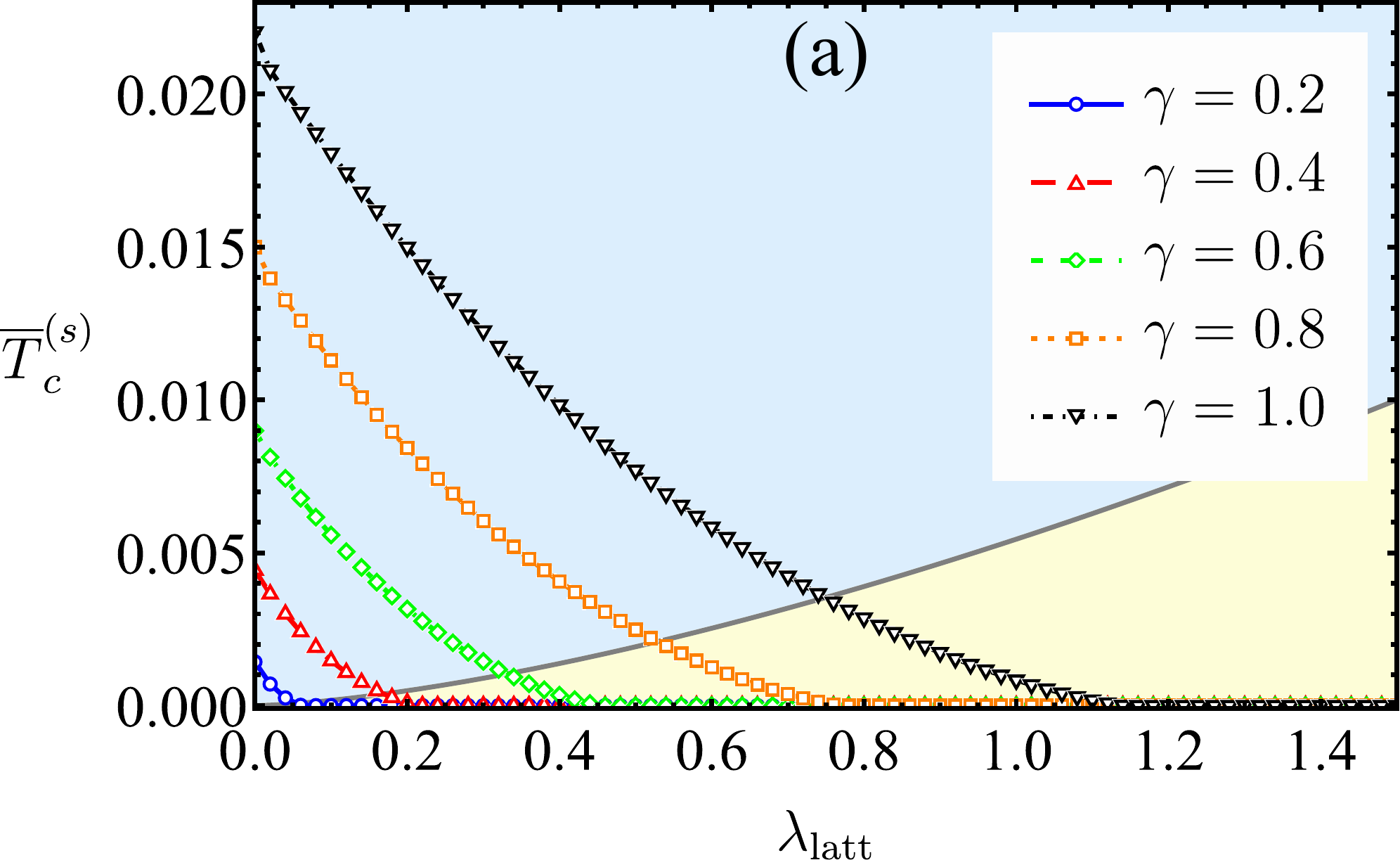} \hfil{} \includegraphics[width=0.45\linewidth,valign=t]{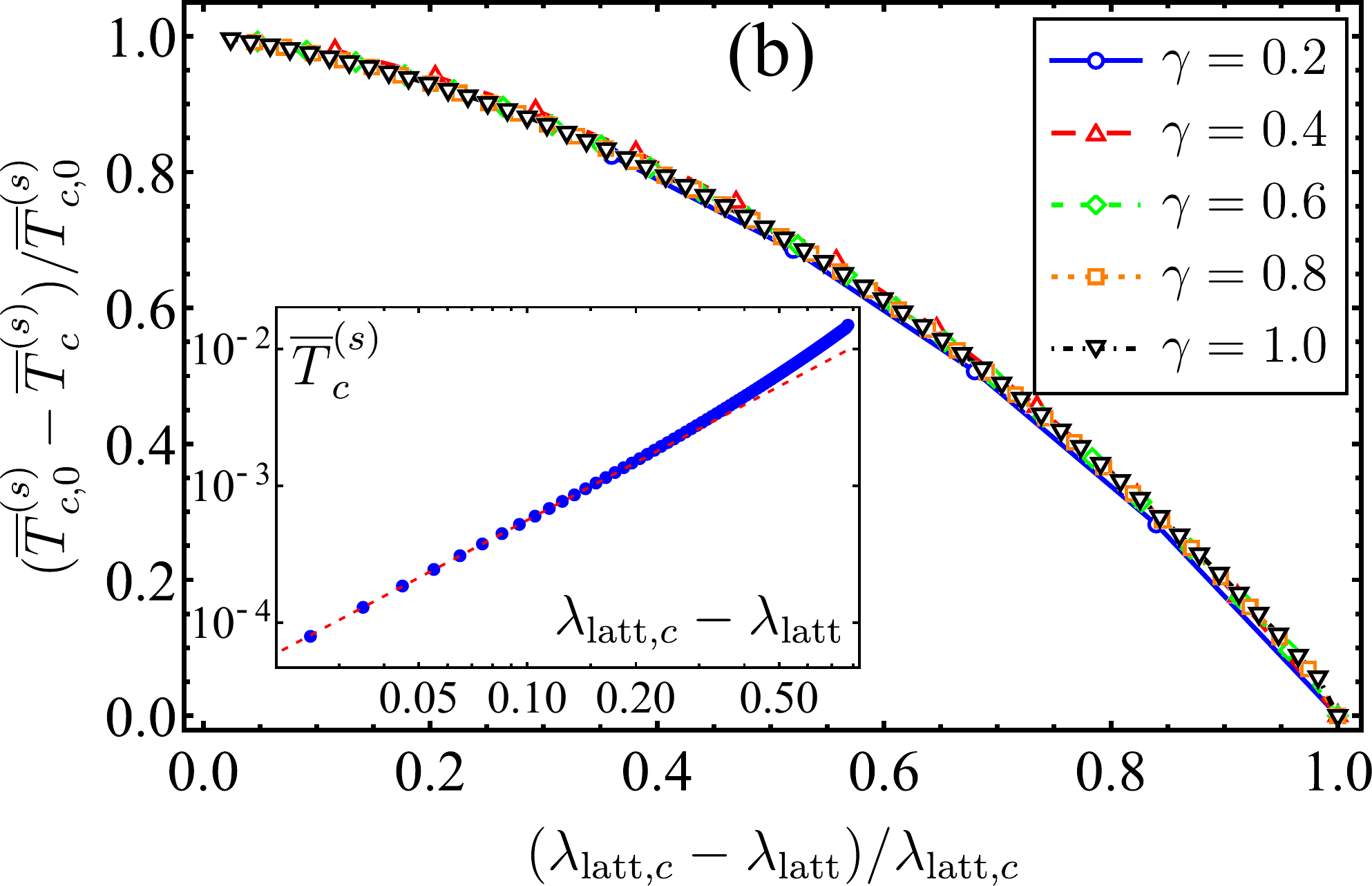} 
\caption{Dependence of the SC critical temperature for \emph{s}-wave pairing on the effective electron interaction $\gamma$, and the nematoelastic coupling $\lambda_\mathrm{latt}$. (a) For a fixed value of $\gamma$, $T^{(s)}_c$ decreases monotonically as $\lambda_\mathrm{latt}$ increases, and becomes zero at a critical coupling $\lambda^{(s)}_{\mathrm{latt}, c} \sim \gamma^{\eta_s}$. The yellow and blue regions refer, respectively, to the emergence of FL and NFL behavior in the absence of SC; they are separated by the crossover (solid) line, which is given by $T_\mathrm{FL} \sim \lambda^{3/2}_\mathrm{latt} E_F$. (b) The behavior of $(\overline{T}^{(s)}_{c, 0} - \overline{T}^{(s)}_c)/\overline{T}^{(s)}_{c, 0}$ as a function of $(\lambda_{\mathrm{latt}, c} - \lambda_\mathrm{latt})/\lambda_{\mathrm{latt}, c}$ is characterized by a universal function, where $\overline{T}^{(s)}_{c, 0} = \overline{T}^{(s)}_{c}(\gamma)$ is the SC critical temperature for \emph{s}-wave pairing in the absence of nematoelastic coupling. The inset shows the power-law dependence of $\overline{T}^{(s)}_c$ on $(\lambda_{\mathrm{latt}, c} - \lambda_\mathrm{latt})$ for $\gamma = 0.8$. The same feature is also found in the transition region for other values of $\gamma$.}\label{Tc_Dependence_Lambda}
\end{figure*}

In Fig. \ref{Tc_Dependence}, we display the dependence of $T_c$ on the effective interaction $\gamma$ for both \emph{s}-wave and \emph{d}-wave symmetries, when $\lambda_\mathrm{latt}$ is held fixed. For $\lambda_\mathrm{latt} = 0$, we find in agreement with Ref. \cite{Klein-PRB(2018)} that both SC critical temperatures for these gap symmetries approach zero as a power-law given by $T^{(s)}_c \sim T^{(d)}_c \propto \gamma^2 E_F$ in the weak coupling regime. Besides, $T^{(s)}_c$ is always larger than $T^{(d)}_c$ and this behavior also occurs in the strong coupling limit, although in this case the difference between $T^{(s)}_c$ and $T^{(d)}_c$ turns out to be more appreciable [see Fig. \ref{Tc_Dependence}(a)]. As $\lambda_\mathrm{latt}$ becomes finite, the system evolves to a SC state only if the electron interaction $\gamma$ is larger than a critical value $\gamma_c$, which according to the results in Figs. \ref{Tc_Dependence}(b) and \ref{Tc_Dependence}(c) increases with $\lambda_\mathrm{latt}$. This bears some resemblance with the $T_c$ behavior of an effective model for low-density systems, when one considers the combined effect of electron-phonon attraction and electron-electron repulsion \cite{Phan-PRB(2022),*Pimenov-arXiv(2021)}. Similarly to what happens with the electrical resistivity of a 2D system described by the present model \cite{Carvalho-PRB(2019),Freire-AP(2020)}, the effect of $\lambda_\mathrm{latt}$ on the SC critical temperature becomes irrelevant in the regime $\gamma \gg \lambda_\mathrm{latt}$. In fact, this can be seen in Fig. \ref{Tc_Dependence}(c), which shows the tendency for the difference $T^{(s)}_c - T^{(d)}_c$ to saturate at a constant, independent of $\lambda_\mathrm{latt}$, as $\gamma$ flows to the strong coupling regime.

We also study the behavior of $T_c$ for \emph{s}-wave and \emph{d}-wave pairing by fixing $\gamma$ and then varying the nematoelastic coupling. As displayed in Fig. \ref{Tc_Dependence_Lambda}(a), the SC critical temperature $T^{(s)}_c$ for \emph{s}-wave pairing decreases monotonically with $\lambda_\mathrm{latt}$ and then becomes zero for nematoelastic couplings above a certain critical value $\lambda_{\mathrm{latt}, c}$. Most importantly, we also find that $T^{(s)}_c$ depends on $(\lambda_{\mathrm{latt}, c} - \lambda_\mathrm{latt})$ as a power-law in the vicinity of $\lambda_{\mathrm{latt}, c}$, as seen in the inset of Fig. \ref{Tc_Dependence_Lambda}(b). We also verified that this dependence does not change, as $\gamma$ varies from the weak to the strong coupling regime. In fact, if we restrict the nemato-elastic coupling to the interval $0 \leqslant \lambda_\mathrm{latt} \leqslant \lambda_{\mathrm{latt}, c}$ [see Fig. \ref{Tc_Dependence_Lambda}(b)], the numerical solution of the Eliashberg equation for $\Delta_n(\theta)$ yields:  
\begin{equation}\label{Eq_SC_Tc}
T^{(\ell)}_{c}(\gamma, \lambda_\mathrm{latt}) = T^{(\ell)}_{c, 0}(\gamma) \bigg[ 1 - \mathscr{F}_{\ell}\bigg( 1 - \frac{\lambda_{\mathrm{latt}}}{\lambda^{(\ell)}_{\mathrm{latt}, c}(\gamma)} \bigg) \bigg],
\end{equation}
where $\ell \in \{s, d \}$, $T^{(\ell)}_{c, 0}(\gamma)$ denotes the SC critical temperature in the absence of nematoelastic coupling, and $\mathscr{F}_{\ell}(x)$ refers to a universal scaling function with the asymptotic dependence
\begin{equation}\label{Eq_Func_Fs}
\mathscr{F}_{\ell}(x) = 
\begin{dcases} 
1 - \mathscr{A}_{\ell} x^{\alpha_\ell}, & \text{if } x \rightarrow 0^{+}, \\
\mathscr{B}_{\ell} (1 - x)^{\beta_\ell}, & \text{if } x \rightarrow 1^{-}.
\end{dcases}
\end{equation}
Figure \ref{Tc_Dependence_Lambda}(b) shows the features of $\mathscr{F}_{\ell}(x)$ for \emph{s}-wave pairing, when one varies its argument from zero to one. In addition, the behavior of $\mathscr{F}_{\ell}(x)$ for \emph{d}-wave pairing is quite similar to the one presented in Fig. \ref{Tc_Dependence_Lambda}(b), although the parameters of that function differ for each gap symmetry. In fact, according to our numerical solution, they evaluate to $\mathscr{A}_{s (d)} \approx 0.68959 \;(0.71402)$, $\mathscr{B}_{s (d)} \approx 1.24779 \;(1.46962)$, $\alpha_{s (d)} \approx 1.42043 \;(1.4689)$, and $\beta_{s (d)} \approx 0.79867 \;(0.872001)$. Although our calculations found that the \emph{s}-wave and \emph{d}-wave $T_c$ are characterized by different exponents $\alpha_\ell$ and $\beta_\ell$, we cannot rule out the possibility that they are degenerate. In fact, since $\alpha_s$ and $\alpha_d$ (or $\beta_s$ and $\beta_d$) are to some extent not far numerically from each other, this particular question can only be completely settled by either working with numerical solutions containing even larger values for the Matsubara cutoff or by finding an exact solution for the Eliashberg equation for $\Delta_n(\theta)$, which are both beyond the scope of the present work. Finally, we also found numerically that the critical nematoelastic coupling has a power-law dependence on $\gamma$ given by $\lambda^{(\ell)}_{\mathrm{latt}, c} \sim \gamma^{\eta_\ell}$, with exponent $\eta_{s (d)} \approx 1.78375 \;(1.32959)$. Consequently, the \emph{s}-wave SC state, when compared to the \emph{d}-wave state, is indeed less susceptible to lattice distortions.

\emph{\textcolor{blue}{Conclusions and outlook.--}} We have studied the emergence of spin-singlet SC with either \emph{s}-wave or \emph{d}-wave symmetry in a 2D electronic system near a nematic QCP, in which the anisotropic nematic fluctuations couple linearly to the elastic modes of the lattice. We find that the main effect of this interaction is to reduce $T_c$ and, for large enough nematoelastic couplings, induce a SC-FL quantum phase transition. This transition, along with the SC-NFL transition that takes place for small values of $\lambda_\mathrm{latt}$, are described by power-laws characterized by distinct critical exponents, which may also differ for \emph{s}-wave and \emph{d}-wave pairings. Lattice effects were shown to play a crucial role on the phase diagram of systems with only a structural QPT to an orthorhombic state \cite{Cowley-PRB(1976),Xu-PRB(2009),Fernandes-PRL(2010),Paul-PRB(2010),Paul-PRB(2017),Schmalian-PRB(2016)}, as is the case of the chemically substituted iron-chalcogenide FeSe$_{1 - x}$S$_x$. Therefore, this compound represents to date the best physical platform to observe the features investigated here. Indeed, it is observed experimentally in the phase diagram of FeSe$_{0.89}$S$_{0.11}$ that $T_c$ suffers a depletion at the onset of nematic order \cite{Coldea-NP(2019)}. According to our present results, this might be related to an increase of the nematoelastic coupling as the nematic QCP is approached. This conclusion is also supported by several recent experimental \cite{Coldea-NP(2019),Bristow-PRR(2020),Gallais-NPJ(2021)} and theoretical works \cite{Carvalho-PRB(2019), Wang-PRB(2019), Freire-AP(2020)} on the behavior of the electrical resistivity exhibited by this correlated material. 

\emph{\textcolor{blue}{Acknowledgments.--}} H.F. acknowledges funding from CNPq under grant numbers 310710/2018-9 and 311428/2021-5. 


%

\end{document}